\documentclass[11pt]{article}
\usepackage{eacl2017}

\usepackage{times}
\usepackage{url}
\usepackage{latexsym}
\usepackage{balance}

\usepackage{graphicx}
\usepackage{multirow}

\usepackage{enumitem}
\setlist{noitemsep}

\usepackage[usenames, dvipsnames]{color}

\eaclfinalcopy % Uncomment this line for the final submission
\title{Character-based Neural Embeddings for Tweet Clustering}

\author{Svitlana Vakulenko \\
  Vienna University of\\ Economics and Business, \\
  MODUL Technology GmbH \\
  {\tt svitlana.vakulenko@wu.ac.at} \\\And
  Lyndon Nixon \\
  MODUL Technology GmbH \\
  {\tt nixon@modultech.eu} \\\And
  Mihai Lupu \\
  TU Wien \\
  Vienna, Austria\\
  {\tt mihai.lupu@tuwien.ac.at} \\}

\date{}

\begin{document}
\maketitle
\begin{abstract}
In this paper we show how the performance of tweet clustering can be improved by leveraging character-based neural networks. The proposed approach overcomes the limitations related to the vocabulary explosion in the word-based models and allows for the seamless processing of the multilingual content. Our evaluation results and code are available on-line\footnote{\url{https://github.com/vendi12/tweet2vec_clustering}}.
\end{abstract}

\section{Introduction}

Our use case scenario, as part of the InVID project\footnote{\url{http://www.invid-project.eu}}, originates from the needs of professional journalists responsible for reporting breaking news in a timely manner. News often appear on social media exclusively or right before they appear in the traditional news media. Social media is also responsible for the rapid propagation of inaccurate or incomplete information (rumors). Therefore, it is important to provide efficient tools to enable journalists rapidly detect breaking news in social media streams \cite{DBLP:conf/icwsm/PetrovicOMMOS13}.

% Computer-assisted information retrieval (IR) systems are designed to help users filter and organize the constant information stream. 

The SNOW 2014 Data Challenge provided the task of extracting newsworthy topics from Twitter. The results of the challenge confirmed that the task is ambitious: The best result was 0.4 F-measure.
% , when evaluated on the topic pool from all the challenge participants. 

Breaking-news detection involves 3 subtasks: selection, clustering, and ranking of tweets. In this paper, we address the task of tweet clustering as one of the pivotal subtasks required to enable effective breaking news detection from Twitter.

% short on related work and method (hypothesis)

Traditional approaches to clustering textual documents involve construction of a document-term matrix, which represents each document as a bag-of-words. These approaches also require language-specific sentence and word tokenization. 

Word-based approaches fall short when applied to social media data, e.g., Twitter, where a lot of infrequent or misspelled words occur within very short documents. Hence, the document representation matrix becomes increasingly sparse.

One way to overcome sparseness in a tweet-term matrix is to consider only the terms that appear frequently across the collection and drop all the infrequent terms. This procedure effectively removes a considerable amount of information content. As a result, all tweets that do not contain any of the frequent terms receive a null-vector representation. These tweets are further ignored by the model and cannot influence clustering outcomes in the subsequent time intervals, where the frequency distribution may change, which hinders the detection of emerging topics.

Artificial neural networks (ANNs) allow to generate dense vector representation (embeddings), which can be efficiently generated on the word- as well as character levels~\cite{DBLP:conf/icml/SantosZ14,DBLP:conf/nips/ZhangZL15,dhingra_tweet2vec:_2016}. The main advantage of the character-based approaches is their language-independence, since they do not require any language-specific parsing.

The major contribution of our work is the  evaluation of the character-based neural embeddings on the tweet clustering task. We show how to employ character-based tweet embeddings for the task of tweet clustering and demonstrate in the experimental evaluation that the proposed approach significantly outperforms the current state-of-the-art in tweet clustering for breaking news detection. 

% 	\item We enhance the ground-truth dataset for breaking news detection task with partial labeling that links tweets to the corresponding topics. This enhancement enables rapid evaluation for tuning the parameters of an algorithm, e.g., distance threshold in the case of hierarchical clustering.
% \end{itemize}

The remaining of this paper is structured as follows: Section~\ref{sec:rel_work} provides an overview of the related work; we describe the setup of an extensive evaluation in Section~\ref{sec:eval}; report and discuss the results in Sections~\ref{sec:results} and \ref{sec:discuss}, respectively; conclusion (Section~\ref{sec:end}) summarizes our findings and directions for future work.

% One way to overcome sparseness in a tweet-term matrix is to consider only the terms that appear frequently across the collection, i.e., document frequency, and drop all the infrequent terms. This procedure effectively removes a considerable amount of information content. As a result, all tweets that do not contain any of the frequent terms receive a 0-vector representation. These tweets are further ignored by the model and can not influence clustering outcomes in the subsequent time intervals, where the frequency distribution may change.

% Artificial neural networks (ANN) allow to generate dense document representations (distributed embeddings). These embeddings accumulate co-occurrence statistics for the terms into condensed vectors of real numbers that efficiently summarize their semantics~\cite{mikolov_distributed_2013}.

% Neural embeddings can be efficiently generated on the character level as well outperforming the word-level baselines on the tasks of language modeling~\cite{DBLP:conf/aaai/KimJSR16}, part-of-speech tagging~\cite{DBLP:conf/icml/SantosZ14}, and text classification\cite{DBLP:conf/nips/ZhangZL15}. Character-based method to process the input string makes this approach language-independent since it does not depend on any language-specific preprocessing procedures, e.g., parsing and tokenization.

\section{Related Work}
\label{sec:rel_work}
\subsection{Breaking news detection}
There has been a continuous effort over the recent years to design effective and efficient algorithms capable of detecting newsworthy topics in the Twitter stream~\cite{hayashi_real-time_2015,ifrim_event_2014,vosecky_dynamic_2013,wurzer_tracking_2015}. These current state-of-the-art approaches build upon the bag-of-words document model, which results in high-dimensional, sparse representations that do not scale well and are not aware of semantic similarities, such as paraphrases. 

The problem becomes evident in case of tweets that contain short texts with a long tail of infrequent slang and misspelled words. The performance of the such approaches over Twitter datasets is very low, with F-measure up to 0.2 against the annotated Wikipidea articles as reference topics~\cite{wurzer_tracking_2015} and 0.4 against the curated topic pool~\cite{papadopoulos_snow_2014}.

\subsection{Neural embeddings} 
Artificial neural networks (ANNs) allow to generate dense vector representations (embeddings). Word2vec~\cite{mikolov_distributed_2013} is by far the most popular approach. It accumulates the co-occurrence statistics of words that efficiently summarizes their semantics. 

Brigadir~et~al.~\shortcite{brigadir_adaptive_2014} demonstrated encouraging results using the word2vec Skip-gram model to generate event timelines from tweets. Moran~et~al.~\shortcite{moran_enhancing_2016} achieved an improvement over the state-of-the-art first story detection (FSD) results by expanding the tweets with their semantically related terms using word2vec.

Neural embeddings can be efficiently generated on the character level as well. They repeatedly outperformed the word-level baselines on the tasks of language modeling~\cite{DBLP:conf/aaai/KimJSR16}, part-of-speech tagging~\cite{DBLP:conf/icml/SantosZ14}, and text classification~\cite{DBLP:conf/nips/ZhangZL15}. The main advantage of the character-based approach is its language-independence, since it does not depend on any language-specific preprocessing.

Dhingra~et~al.~\shortcite{dhingra_tweet2vec:_2016} proposed training a recurrent neural network on the task of hashtag prediction. Vosoughi~et~al.~\shortcite{vosoughi_tweet2vec:_2016} demonstrated an improved performance of a character-based neural autoencoder on the task of paraphrase and semantic similarity detection in tweets.

Our work extends the evaluation of the Tweet2Vec model~\cite{dhingra_tweet2vec:_2016} to the tweet clustering task, versus the traditional document-term matrix representation. To the best of our knowledge, this work is the first attempt to evaluate the performance of character-based neural embeddings on the tweet clustering task.

\section{Experimental Evaluation}
\label{sec:eval}
\subsection{Dataset}
\paragraph{Description and preprocessing.}
We use the SNOW 2014 test dataset~\cite{papadopoulos_snow_2014} in our evaluation. It contains the IDs of about 1 million tweets produced within 24 hours. 

We retrieved 845,626 tweets from the Twitter API, since other tweets had already been deleted from the platform. The preprocessing procedure: remove RT prefixes, urls and user mentions, bring all characters to lower case and separate punctuation with spaces (the later is necessary only for the word-level baseline).

The dataset is further separated into 5 subsets corresponding to the 1-hour time intervals (18:00, 22:00, 23:15, 01:00 and 01:30) that are annotated with the list of breaking news topics. In total, we have 48,399 tweets for clustering evaluation; the majority of them (42,758 tweets) are in English.

% Only 26\% (222,413) of these tweets contained hashtags and could therefore be used for training the model. 
% For comparability of results, the preprocessing procedure was the same as that of Dhingra et al.~\cite{dhingra_tweet2vec:_2016}: remove RT prefixes and hashtags, and substitute user mentions and URLs with placeholders. In this way we obtained 88,148 unique tweets (10\% of the original dataset) that we used for training the Tweet2Vec model. 
% See Table~\ref{table:repr_power} for the dataset statistics.

The dataset comes with the list of the breaking news topics. These topics were manually selected by the independent evaluators from the topic pool collected from all challenge participants (external topics).  The list of topics contains 70 breaking news headlines extracted from tweets (e.g., ``The new, full Godzilla trailer has roared online''). Each topic is annotated with a few (at most 4) tweet IDs, which is not sufficient for an adequate evaluation of a tweet clustering algorithm. 

\paragraph{Dataset extension.}
We enrich the topic annotations by collecting larger tweet clusters using fuzzy string matching\footnote{\url{https://github.com/seatgeek/fuzzywuzzy}} for each of the topic labels. Fuzzy string matching uses the Levenstein (edit) distance~\cite{levenshtein1966binary} between the two input strings as the measure of similarity. Levenstein distance corresponds to the minimum number of character edits (insertions, deletions, or substitutions) required to transform one string into the other. We choose only the tweets for which the similarity ratio with the topic string is greater than 0.9 threshold. 

A sample tweet cluster produced with the fuzzy string matching for the topic ``Justin Trudeau apologizes for Ukraine joke'': 
\begin{itemize}
	\item Justin Trudeau apologizes for \textit{Ukraine joke}: Justin Trudeau said he's spoken the head...% of the Ukrainian...%\url{http://t.co/PQE5fsP0Sd} %\#thestar
	\item Justin Trudeau apologizes for \textit{Ukraine comments} \url{http://t.co/7ImWTRONXt}
    \item Justin Trudeau apologizes for \textit{Ukraine hockey joke} \#cdnpoli% \url{http://t.co/onT162CKPt} %\url{http://t.co/38yw25RgfX}
\end{itemize}

In total, we matched 2,585 tweets to 132 clusters using this approach. The resulting tweet clusters represent the ground-truth topics within different time intervals. The cluster size varies from 1 to 361 tweets with an average of 20 tweets per cluster (median: 6.5). 

This simple procedure allows us to automatically generate high-quality partial labeling. We further use this topic assignment as the ground-truth class labels to automatically evaluate different flat clustering partitions.

\subsection{Tweet representation approaches}

\paragraph{TweetTerm.}
Our baseline is the tweet representation approach that was used in the winner-system of SNOW 2014 Data Challenge\footnote{\url{https://github.com/heerme/twitter-topics}}~\cite{ifrim_event_2014}. This approach represents a collection of tweets as a tweet-term matrix by keeping the bigrams and trigrams that occur at least in 10 tweets. 

% All user mentions were discarded in the preprocessing stage, but hashtags were preserved. 
% Further on we refer to this baseline approach as \textbf{TweetTerm} for brevity.

\paragraph{Tweet2Vec.}

% We propose to replace the document-term matrix with the character-based embeddings to represent a corpus of tweets and 	use these representations for clustering. The approach involves two major steps: (1) training a neural network on the character level to encode tweets optimized for the task of predicting the previously removed hashtags; (2) hierarchical and flat clustering of the produced tweet embeddings. These clusters are then used as the basis for identification of newsworthy events (topics). 

% A neural embedding is a dense vector representation produced with a neural network that efficiently stores co-occurrence statistics as real-valued vectors. We used Tweet2Vec to train a neural network able to produce high-quality tweet embeddings. 

This approach includes two stages: (1) \textit{training} a neural network to predict hashtags using the subset of tweets that contain hashtags (88,148 tweets  in our case); (2) \textit{encoding}: use the trained model to produce tweet embeddings for all the tweets regardless whether they contain hashtags or not. We use Tweet2Vec implementation\footnote{\url{https://github.com/bdhingra/tweet2vec}} to produce tweet embeddings.

% Each character is encoded as a one-hot vector of size equal to the number of unique characters (alphabet).

Tweet2Vec is a bi-directional recurrent neural network that consumes textual input as a sequence of characters. The network architecture includes two Gated Recurrent Units (GRUs)~\cite {DBLP:conf/emnlp/ChoMGBBSB14}: forward and backward GRUs. GRU is an optimized version of a Long Short-Term Memory\hyphenation{Me-mo-ry} (LSTM) architecture~\cite{DBLP:journals/neco/HochreiterS97}. It includes 2 gates that control the information flow. The gates (reset and update gate) regulate how much the previous output state ($h_{t-1}$) influences the current state ($h_{t}$).

The two GRUs are identical, but the backward GRU receives the same sequence of tweet-characters in reverse order. Each GRU computes its own vector-representation for every substring ($h_{t}$) using the current character vector ($x_t$) and the vector-representation it computed a step before ($h_{t-1}$). These two representations of the same tweet are combined in the next layer of the neural network to produce the final tweet embedding
(see more details in Dhingra et.al.~\shortcite{dhingra_tweet2vec:_2016}). 

The network is trained in minibatches with an objective function to predict the previously removed hashtags. A hashtag can be considered as the ground-truth cluster label for tweets. Therefore, the network is trained to optimize for the correct tweet classification, which corresponds to a supervised version of the tweet clustering task annotated with the cluster assignment, i.e. hashtags. 

In order to predict the hashtags the tweet embeddings are passed through the linear layer, which produces the output in the size of the number of hashtags, which we observed in the training dataset. The softmax layer on top normalizes the scores from the linear layer to generate the hashtag probabilities for every input tweet. 

% \textcolor{red}{if we have space, perhaps add a diagram of the network}

% The number of hidden states, which corresponds to the dimensionality of the produced tweet-vectors, is one of the hyper-parameters of the model that has to be set before training the model. Dhingra~et.~al.~\cite{dhingra_tweet2vec:_2016} reports that the models with 300-500 hidden states provide similar performance results on the hashtag prediction task.

Tweet embeddings are produced by passing the tweets through the trained Tweet2Vec model (encoder). In this way we can obtain vector representations for all the tweets including the ones that do not contain any hashtags. The result is a matrix of size $n \times h$, where $n$ is the number of tweets and $h$ is the number of hidden states (500).

% We use the \textit{Tweet2Vec} implementation
% with the default parameters.\textcolor{red}{could we say why the default parameters is a sensible choice?}

\subsection{Clustering} 
To cluster tweet vectors  (character-based tweet embeddings produced by the neural network for Tweet2Vec evaluation or the document-term matrix for TweetTerm) we employ the hierarchical clustering algorithm implementation from \textit{fastcluster} library~\cite{JSSv053i09}.

Hierarchical clustering includes computing pairwise distances between the tweet vectors, followed by their linkage into a single dendrogram.  There are several distance metrics (Euclidean, Manhattan, cosine, etc.) and linkage methods to compare distances (single, average, complete, weighted, etc.). We evaluated the performance of different methods using the cophenetic correlation coefficient~(CPCC)~\cite{sokal1962comparison} and found the best performing combination: Euclidean distance and average linkage method.

The hierarchical clustering dendrogram can produce $n$ different flat clusterings for the same dataset: from \textit{n} single-member clusters with one document per cluster to a single cluster that contains all \textit{n} documents. The distance threshold defines the granularity (number and size) of the produced clusters. 

\subsection{Distance threshold selection} 

Grid search helps us to determine the optimal distance threshold for the dendrogram cut-off. We generated a list of values in the range from 0.1 to 1.5 with 0.1 increment step and examine their performance with respect to the ground-truth cluster assignment. We produce flat clusterings for each value of the distance threshold from the grid and compare them with respect to the quality metrics.

Since we also want to be able to select the optimal distance threshold in absence of the true labels, we examine the scores provided by the mean Silhouette coefficient~\cite{ROUSSEEUW198753}. Silhouette is an unsupervised intrinsic evaluation metric (cluster validity index) that measures the quality of the produced clusters and can be used for unsupervised intrinsic evaluation (i.e., without the ground-truth labels). It was reported to outperform alternative methods in a comparative study of 30 validity indices~\cite{DBLP:journals/pr/ArbelaitzGMPP13}.

% We use the Silhouette coefficient~\cite{ROUSSEEUW198753} to find an optimal value for the \textbf{distance threshold} in a hierarchical clustering dendrogram. Silhouette is a cluster validity index that measures the quality of the produced clusters and can be used for unsupervised intrinsic evaluation (i.e., without the ground-truth labels). 

% We use Silhouette because Arbelaitz~et~al.~\cite{DBLP:journals/pr/ArbelaitzGMPP13} conducted an extensive study of 30 different cluster validity indices and compared their performance using three clustering algorithms: k-means, Ward, and Average-linkage. They reported that there was no single index that outperformed the rest in all the evaluation settings, but the Silhouette coefficient obtained the best results in many of them including hierarchical algorithms.

\subsection{Clustering Evaluation Metrics}
We evaluate the clustering results using the standard metrics for extrinsic clustering evaluation: homogeneity, completeness, V-Measure~\cite{DBLP:conf/emnlp/RosenbergH07}, Adjusted Rand Index (ARI)~\cite{hubert1985comparing} and Adjusted Mutual Information (AMI)~\cite{DBLP:journals/jmlr/NguyenEB10}. All metrics return a score on the range [0; 1] for the pair of sets that contain ground truth and cluster labels as input. The higher the score the more similar the two clusterings are.

The \textbf{Homogeneity} score represents the measure for purity of the produced clusters. It penalizes clustering, where members of different classes get clustered together. Thus, the best homogeneity scores are always at the bottom of the dendrogram, i.e., at the level of the leaves, where each document belongs to its own cluster. \textbf{Completeness}, on the contrary, favors larger clusters and reduces the score if the members of the same class are split into different clusters. Therefore, the top of the dendrogram, where all the documents reside in a single cluster always achieves the maximum completeness score.

\textbf{V-Measure} is designed to balance out the two extremes of homogeneity and completeness. It is the harmonic mean of the two and corresponds to the Normalized Mutual Information (NMI) score. 

\textbf{AMI} score is an extension of NMI adjusted for chance. The more clusters are considered the more chance the labelings correlate. AMI allows us to compare the clustering performance across different time intervals since it normalizes the score by the number of labeled clusters in each interval. 

Finally, \textbf{ARI} is an alternative way to assess the agreement between two clusterings. It counts all pairs clustered together or separated in different clusters. ARI also accounts for the chance of an overlap in a random label assignment.

\begin{table*}[h]
\centering
\resizebox{\textwidth}{!}{%
\begin{tabular}{|c|c|l|l|l|l|l|l|c|c|c|}
\hline
\textbf{Interval} & \textbf{Tweets} & \textbf{Model} & \textbf{Dimensions} & \textbf{Distance threshold} & \textbf{Clusters} & \textbf{Homogeneity} & \textbf{Completeness} & \textbf{V-Measure} & \textbf{ARI} & \textbf{AMI} \\ \hline
\multirow{2}{*}{18:00} & \multirow{2}{*}{10,344} & Tweet2Vec & 500 & 1 & 3026 & \textbf{0.9958} & 0.9453 & \textbf{0.9699} & \textbf{0.9804} & \textbf{0.9376} \\ \cline{3-11} 
 &  & {TweetTerm} & 433 & 1-1.3 & 66-79 & 0.9277 & \textbf{1} & 0.9625 & 0.949 & 0.9216 \\ \hline
 \hline
\multirow{2}{*}{22:00} & \multirow{2}{*}{14,471} & Tweet2Vec & 500 & 0.9 & 5292 & \textbf{1} & 0.9601 & \textbf{0.9796} & \textbf{0.9922} & \textbf{0.9571} \\ \cline{3-11} 
 &  & {TweetTerm} & 589 & 0.7-1.3 & 93-118 & 0.9385 & \textbf{0.9969} & 0.9668 & 0.9859 & 0.9359 \\ \hline
 \hline
\multirow{2}{*}{23:15} & \multirow{2}{*}{8,231} & Tweet2Vec & 500 & 0.8 & 3986 & \textbf{1} & 0.98 & \textbf{0.9899} & \textbf{0.9948} & \textbf{0.9743} \\ \cline{3-11} 
 &  & {TweetTerm} & 565 & 0.01-1.3 & 67-142 & 0.8062 & \textbf{0.9978} & 0.8918 & 0.7344 & 0.7763 \\ \hline
 \hline
\multirow{2}{*}{01:00} & \multirow{2}{*}{5,123} & Tweet2Vec & 500 & 0.9 & 2242 & \textbf{1} & 0.8877 & \textbf{0.9405} & \textbf{0.8668} & \textbf{0.8327} \\ \cline{3-11} 
 &  & {TweetTerm} & 721 & 0.8-1.3 & 71-111 & 0.8104 & \textbf{1} & 0.8953 & 0.8188 & 0.7666 \\ \hline
 \hline
\multirow{2}{*}{01:30} & \multirow{2}{*}{4,589} & Tweet2Vec & 500 & 0.9 & 2091 & \textbf{1} & 0.8762 & \textbf{0.934} & \textbf{0.8089} & \textbf{0.8129} \\ \cline{3-11} 
 &  & {TweetTerm} & 635 & 1.2-1.3 & 64-78 & 0.8024 & \textbf{1} & 0.8903 & 0.7809 & 0.754 \\ \hline
\end{tabular}%
}
\caption{Results of clustering evaluation on the English-language dataset}
\label{table:results}
\end{table*}

\subsection{Manual Cluster Evaluation}
Our partial labeling covers a small subset of the data and by design provides the clusters with the high degree of string overlap with the annotated topics. Therefore, we extend the clustering evaluation to the rest of the dataset to evaluate whether the models can uncover less straight-forward semantic similarities in tweets. We select the results for manual evaluation motivated by the cluster label (headline) selection task.

The next step in the breaking news detection pipeline after the clustering task is headline selection (cluster labeling task). The most common approach to label a cluster of tweets is to select a single tweet as a representative member for the whole cluster \cite{papadopoulos_snow_2014}. We decided to test this assumption and manually check how many clusters loose their semantics when represented with a single tweet. 

Headline selection motivates the coherence assessment of the produced clusters since the clusters discarded at this stage will never make it to the final results. To explore coherence of the produced clusters we pick several tweets in each cluster and check whether they are semantically similar.

The tweet selected as a headline (cluster label) can be the first published tweet as in First Story Detection (FSD) task, also used in Ifrim~et~al~\shortcite{ifrim_event_2014}. Alternative approaches include selection of the most recent tweet published on the topic, or the tweet that is semantically most similar to all other tweets in the cluster, i.e., the tweet closest to the centroid of the cluster (\textbf{medoid-tweets}). Therefore, we sample 5 tweets from each cluster: the first published tweet, the most recent tweet and three medoid-tweets.

We set up a manual evaluation task as follows:

\begin{enumerate}
	\item Take the top 20 largest clusters sorted by the number of tweets that belong to the cluster.
    \item For each cluster:
    \begin{enumerate}
      \item Take the first and the last published tweet (tweets are previously sorted by the publication date).
      \item Take three medoid-tweets, i.e., the tweets that appear closest to the centroid of the cluster. 
      \item Add the 5 tweets to the set associated with the cluster (removing exact duplicate tweets) 
    \end{enumerate}
    \item For all clusters, where the set of selected tweets contains at least two unique tweets: 4 human evaluators independently assess the coherence of each cluster.  
\end{enumerate}

According to the evaluation setup each model produced 20 top-clusters for each of the 5 intervals, i.e., $20 \times 5 = 100$ clusters per model. We manually evaluate only the clusters that contain more than 1 distinct representative tweet  (\textbf{Clusters$>$1}). All other clusters, i.e., the ones for which all 5 selected tweets are identical (\textbf{Clusters=1}), are considered correct by default.

Results for all 5 intervals were evaluated together in a single pool and the models were anonymized to avoid biases. Each evaluator independently assigned a single score to each cluster:
\begin{itemize}
	\item \textbf{Correct} -- all tweets report the same news;
    \item \textbf{Partial} -- some tweets are not related;
    \item \textbf{Incorrect} -- all tweets are not related.
\end{itemize}

Partial and Incorrect labels reflect different types of clustering errors. Partial error is less severe indicating that the tweets of the cluster are semantically similar, but they report different news (events) and should be split into several clusters. Incorrect clusters indicate a random collection of tweets with no semantic similarities.

\begin{table*}[h]
\centering
% \resizebox{\textwidth}{!}{%
\begin{tabular}{|l|l|c|c|c|c|c|c|}
\hline
\multicolumn{1}{|c|}{\multirow{3}{*}{Model}} & \multicolumn{1}{c|}{\multirow{3}{*}{Dataset}} & \multirow{3}{*}{Clusters} & \multicolumn{3}{c|}{Correct (\%)} & \multicolumn{2}{c|}{Errors (\%)} \\ \cline{4-8} 
\multicolumn{1}{|c|}{} & \multicolumn{1}{c|}{} &  & Clusters=1 & Clusters\textgreater1 & Total & Partial & Incorrect \\ \hline
% \multicolumn{1}{|c|}{} & \multicolumn{1}{c|}{} &  & \% & \% & \% & \% & \% \\ \hline
Tweet2Vec & English & 100 & \textbf{80} & 8.3 & \textbf{88.3} & 10 & \textbf{1.8} \\ \hline
% TweetTerm & English & 100 & 67 & 16.5 & 83.5 & 8.5 & 8 \\ \hline
TweetTerm  & English & 95 & 71 & \textbf{17.4} & 87.9 & 8.9 & 3.2 \\ \hline
\hline
Tweet2Vec & Multilingual & 100 & 67 & 12.5 & \textbf{79.5} & 13 & 7.5\\ \hline
\end{tabular}
\caption{Results of manual cluster evaluation. Note: the last row shows results on a different dataset and can not be directly compared with the other models.}
\label{table:maneval}
% }
\end{table*}

\section{Results}
\label{sec:results}
\subsection{Results of Clustering Evaluation}

Table~\ref{table:results} summarizes the results of our evaluation using the ground-truth partial labeling. The scores highlighted with the bold font indicate the best result among the two competing approaches for the same subset of tweets corresponding to the respective time interval. 

Tweet2Vec exhibits better clustering performance comparing to the baseline according to the majority of the evaluation metrics in all the intervals. In all cases Tweet2Vec model wins in terms of Homogeneity score and TweetTerm wins in Completeness. This result shows that Tweet2Vec is better at separating tweets that are not similar enough than the baseline model. Tweet2Vec fails only once to perfectly separate the ground-truth clusters (18:00 interval). This result shows that Tweet2Vec is able to replicate the results of the fuzzy string matching algorithm that was used to generate the ground-truth labeling.

\subsection{Results of Distance Threshold Selection}
% Silhouette coefficient does not correlate with the V-Measure score. It gives preference to the more fine-grained clustering . 

\begin{figure}
\centering
\includegraphics[width=0.5\textwidth]{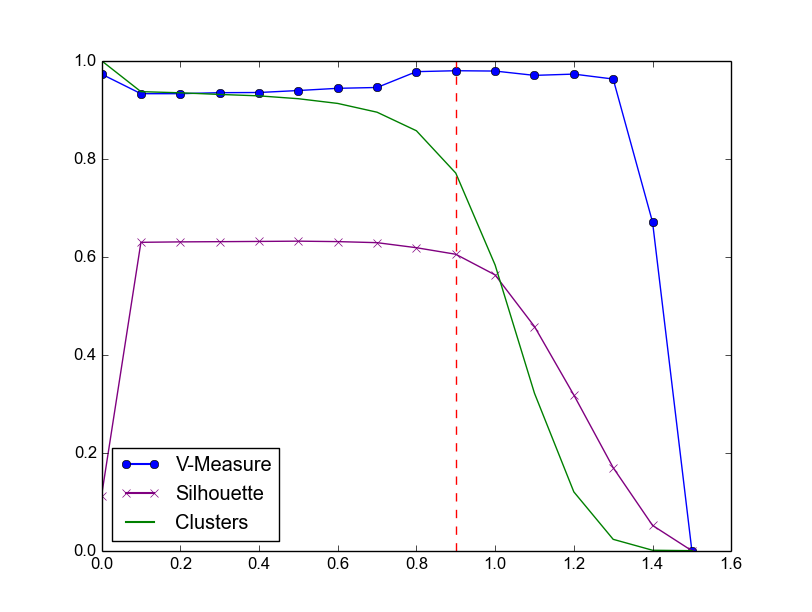}
\caption{Correlation between the V-Measure, Silhouette coefficient and the number of clusters per tweet (Tweet2Vec 22:00 interval). The vertical red line indicates the maximum V-Measure score.}
\label{fig:sil}
\end{figure}

The rise in V-Measure correlates with the decline of the Silhouette coefficient and the steep drop in the number of produced clusters (see Figure~\ref{fig:sil}). We observed that the optimal distance threshold for Tweet2Vec clustering according to V-Measure is on the interval [0.8; 1] (see Table~\ref{table:results}: Distance threshold), which is also consistent with the findings reported in Ifrim et. al~\shortcite{ifrim_event_2014}. 

% However, there is no evidence that this pattern will generalize outside of this dataset.

\subsection{Results of Manual Cluster Evaluation}

% Please add the following required packages to your document preamble:
% \usepackage[normalem]{ulem}
% \useunder{\uline}{\ul}{}

Results of the manual cluster evaluation by four independent evaluators are summarized in Table~\ref{table:maneval}. Bold font indicates the maximum scores achieved across the competing representation approaches. Tables~\ref{table:Tweet2Vec} and \ref{table:TweetTerm} show sample clusters produced by both models alongside their average score.

TweetTerm assigns a 0-vector representation to tweets that do not contain any of the frequent terms. Hence, all these tweets end up in a single ``garbage'' cluster. Therefore, we discount the number of the expected ``garbage'' clusters (1 cluster per interval = 5 clusters) from the score count for TweetTerm (Table~\ref{table:maneval}).

Tweet2Vec model produces the largest number of perfectly homogeneous clusters for which all 5 selected tweets are identical (see Table~\ref{table:maneval} column Clusters=1). The percentage of correct results among the manually evaluated clusters is higher for the TweetTerm model, but the number of errors (Incorrect) is higher as well. Tweet2Vec produced the highest total \% of correct clusters due to the larger proportion of detected clusters that contain identical tweets (Clusters=1). Tweet2Vec also produced the least number of incorrect clusters: at most 2 incorrect clusters per 100 clusters (Precision: 0.98).
%put into motivation: TweetTerm cannot generate a distinct vector representation for every distinct tweet in the dataset due to the inherent limitations of the model.

The results of Tweet2Vec on the multilingual dataset are lower than on the English-language tweets. However, we do not have alternative results to compare since the baseline approach is not language-independent and requires additional functionality (word-level tokenizers) to handle tweets in other languages, e.g., Arabic or Chinese. We provide this evaluation results to demonstrate that Tweet2Vec overcomes this limitation and is able to cluster tweets in different languages. In particular, we obtained correct clusters of Russian and Arabic tweets.

We observed that leaving the urls does not significantly affect clustering performance, i.e., the model tolerates noise. However, replacement of the urls and user mentions with placeholders as in Dhingra~et.~al.~\shortcite{dhingra_tweet2vec:_2016} generates syntactic patterns in text, such as @user @user @user, which causes semantically unrelated tweets appear within the same cluster.

\begin{table*}[h]
\centering
\small
\begin{tabular}{|l|l|}
\hline
Sample Cluster & Evaluation \\ \hline
\begin{tabular}[c]{@{}l@{}}video : \textbf{bitcoin} \textbf{:} \textbf{mtgox} exchange goes offline - \textbf{bitcoin} , a virtual currency ...\\ the slow-motion collapse of \textbf{mt . gox} is \textbf{bitcoin}'s first financial crisis \textbf{:} now \textbf{bitcoin} users ...\\ Disastro \textbf{bitcoin} : \textbf{mt . gox} cessa ogni attivite ... \textbf{:} \textbf{mt . gox} , il pi ù grande cambiavalute \textbf{bitco} ...\end{tabular} & Correct \\ \hline
% \begin{tabular}[c]{@{}l@{}}энергетические браслеты с турмалином в украине\\ энергетические браслеты купить в украине через интернет магазин\\ энергетические браслеты в украине январь 2014 ярославль\end{tabular} & Correct \\ \hline
% \begin{tabular}[c]{@{}l@{}}syria alarabiya aljazeera تحليق الطيران الحربي فوق مدينة عدرا العمالية\\ syria joradn saudi تحليق الطيران الحربي فوق مدينة عدرا العمالية\\ syria ksa oman تحليق الطيران الحربي فوق مدينة عدرا العمالية\end{tabular} & Correct \\ \hline
\begin{tabular}[c]{@{}l@{}}\textbf{california couple finds} time capsules \textbf{worth} \textbf{\$10 million}\\ \textbf{californian couple finds} \textbf{\$10 million} \textbf{worth} of gold coins in tin can\end{tabular} & Correct \\ \hline
\begin{tabular}[c]{@{}l@{}}\textbf{ukraine} puts off vote on new government despite eu pleas for quick action \textbf{-} washington post ...\\ \textbf{ukraine} truce shattered , death toll hits 67 \textbf{-} kiev (reuters) \textbf{-} ukraine suffered its bloodiest day ...\\ \textbf{ukraine} fighting leaves at least 18 dead as kiev barricades burn \textbf{-} clashes in ukraine ...\end{tabular} & Partial \\ \hline
% \begin{tabular}[c]{@{}l@{}}"today, i am joined by researchers who invent some of the most advanced metals on the planet ." president obama\\ "we now have 4 million americans who have signed up for quality private health insurance through the marketplace ." president obama\end{tabular} & Partial \\ \hline
\begin{tabular}[c]{@{}l@{}}\textbf{are you} going to come on his network and get poor ratings too \textbf{?}\\ \textbf{are you} sold on the waffle taco \textbf{?}\end{tabular} & Incorrect \\ \hline
\begin{tabular}[c]{@{}l@{}}\textbf{the} chromecast app flood has started \textbf{by}\\ \textbf{the} importance of emotion in design \textbf{by}\end{tabular} & Incorrect \\ \hline
\end{tabular}
\caption{Tweet2Vec sample results. Rows of the table show sample tweet clusters. Each line within the row corresponds to a separate tweet (after preprocessing, i.e. usernames and urls removed.)}
\label{table:Tweet2Vec}
\end{table*}

\begin{table*}[h]
\centering
\small
\begin{tabular}{|l|l|}
\hline
Sample Cluster & Evaluation \\ \hline
\begin{tabular}[c]{@{}l@{}}obama : michelle and i were saddened to hear of the passing of \textbf{harold ramis}...\\ touching tribute to ghostbusters star \textbf{harold ramis} from comic artist\\ on the joyful comedy of \textbf{harold ramis}\end{tabular} & Correct \\ \hline
\begin{tabular}[c]{@{}l@{}}major tokyo-based bitcoin exchange \textbf{mt . gox goes dark}\\ "bitcoin exchange giant \textbf{mt . gox goes dark} | popular science "\end{tabular} & Correct \\ \hline
\begin{tabular}[c]{@{}l@{}}\textbf{obesity rate for young children} plummets \textbf{43 \%} in a \textbf{decade}\\ the national \textbf{obesity rate for young children} dropped \textbf{43 \%} over the past \textbf{decade}\end{tabular} & Correct \\ \hline
% \begin{tabular}[c]{@{}l@{}}ukraine's acting leader still seeking consensus on interim government - los angeles times news\\ news \textbf{ukraine} not east-west fight , says us world usa\\ bbc \textbf{ukraine} not east-west fight , says us\end{tabular} & Partial \\ \hline
\begin{tabular}[c]{@{}l@{}}diplomatic pressure is unlikely to reverse uganda's cruel \textbf{anti-gay} law\\ provisions of arizona proposed \textbf{anti-gay} law\\ even mitt romney wants arizona's governor to veto the state's \textbf{anti-gay} bill\\ icymi : arizona pizzeria response to state \textbf{anti-gay} bill\end{tabular} & Partial \\ \hline
% \begin{tabular}[c]{@{}l@{}}today i had a beautiful surprise , i was able to facetime with my idol \textbf{muhammad ali}\\ experience the fight that made \textbf{muhammad ali} a champion\end{tabular} & Incorrect \\ \hline
\begin{tabular}[c]{@{}l@{}}amazing debate nic ! \textbf{well done} !\\ \textbf{well done} 4 -0\\ \textbf{well done} ! i find running so difficult . feel proud !\\ \textbf{well done} him :-)\\ \textbf{well done} nicola my money is on you you done it well tonight ??\end{tabular} & Incorrect \\ \hline
\end{tabular}
\caption{TweetTerm sample results. Rows of the table show sample tweet clusters.}
% Each line within the row corresponds to a separate tweet (after preprocessing, i.e. usernames and urls removed.)}
\label{table:TweetTerm}
\end{table*}

\section{Discussion}
\label{sec:discuss}
Our experimental evaluation showed that the character\hyphenation{cha-ra-cter}-based embeddings produced with a neural network outperform the document-term baseline on the tweet clustering task.  The baseline approach (TweetTerm) shows a very good performance in comparison with the simplicity of its implementation, but it naturally falls short in recognizing patterns beyond simple n-gram matching.

We attribute this result to the inherent limitation of the document-term model retaining only the frequent terms and disregarding the long tail of infrequent patterns. This limitation appears crucial in the task of emergent news detection, in which the topics need to be detected long before they become popular. Neural embeddings, in contrast, can retain a sufficient level of detail in their representations and are able to mirror the fuzzy string matching performance beyond simple n-gram matching. 

It becomes apparent from the sample clustering results (Tables~\ref{table:Tweet2Vec} and \ref{table:TweetTerm}) that both models perform essentially the same task of unveiling patterns shared between a group of strings. While TweetTerm operates only on the patterns of identical n-grams, Tweet2Vec goes beyond this limitation by providing room for a variation within the n-gram substring similar to fuzzy string matching. This effect allows to capture subtle variations in strings, e.g., misspellings, which word-based approaches are incapable of.

Our error analysis also revealed the limitation of the neural embeddings to distinguish between semantic and syntactic similarity in strings (see Incorrect samples in Table~\ref{table:Tweet2Vec}). Tweet2Vec, as a recurrent neural network approach, represents not only the characters but also their order in string that may be a false similarity signal. It is evident that the neural representations in our example would benefit from the stop-word removal or an analogous to TF/IDF weighting scheme to avoid capturing punctuation and other merely syntactic patterns.

% However, we did not find evidence in our results that the model is able to capture synonymity and paraphrases beyond fuzzy string matching and hashtag prediction.

% \textbf{Languages.} 
% The model tends to separate tweets written in different languages into different clusters. Therefore, there is no evidence we can achieve cross-lingual effect, i.e. capturing semantic similarity across languages and detecting similar reports written in different languages.

% \textbf{Syntactic patterns.} The model captures template-based patterns in text very well, such as: \textcolor{red}{i don't understand the patters below. Templates are never mentioned before - what are they and where do they come from? }

% 2014 02 26 10 30 35 panamabitcoins ticker is 562.09 balboas

% 2014 02 26 09 20 29 panamabitcoins ticker is 566.35 balboas

% android app for viewing espresotv 26.02 12 30 01

% android app for viewing espresotv 25.02 21 15 01

% This result suggests that the model can be adopted to detect automatically generated template-based texts (bots).

% \textbf{Model update and scalability.} We also indicate several challenges related to the practical implementation of the proposed approach. The model has to be regularly updated in order to reflect semantic drifts in the information space. Also, scalability of the neural-network-based approach was not evaluated. However, this requirement is one of the essentials for the real-time breaking news detection system.

\paragraph{Limitations.} Neural networks gain performance when more data is available. We could use only 88,148 tweets from the dataset to train the neural network, which can appear insufficient to unfold the potential of the model to recognize more complex patterns. Also, due to the scarce annotation available we could use only a small subset of the original dataset for our clustering evaluation. Since most of the SNOW tweets are in English, another dataset is needed for comprehensive multilingual clustering evaluation.

% On the other hand, the model trained to predict hashtags may not be powerful enough (too coarse-grained) to capture more subtle text similarities. An alternative architecture, such as variational autoencoder~\cite{vosoughi_tweet2vec:_2016}, may be more appropriate. Unfortunately training this model is very computationally expensive (more than a week) and is not easy to scale to real-time stream processing.

\section{Conclusion}
\label{sec:end}
We showed that character-based neural embeddings enable accurate tweet clustering with minimum supervision. They provide fine-grained representations that can help to uncover fuzzy similarities in strings beyond simple n-gram matching. We also demonstrated the limitation of the current approach unable to distinguish semantic from syntactic patterns in strings, which provides a clear direction for the future work.

% Future work
% In future work we plan to work on adding relevance assessment and ranking mechanisms on top of the clustering results to estimate newsworthiness of the clusters and increase precision for the task of breaking news detection. Furthermore, timeliness and scalability of the proposed approach has to be evaluated.
\section{Acknowledgments}
The presented work was supported by the InVID Project (http://www.invid-project.eu/), funded by the European Union's Horizon 2020 research and innovation programme under grant agreement No 687786. Mihai Lupu was supported by Self-Optimizer (FFG 852624) in the EUROSTARS programme, funded by EUREKA, the BMWFW and the European Union, and ADMIRE (P25905-N23) by FWF. We thank to Bhuwan Dhingra for the support in using Tweet2Vec and Linda Andersson for the review and helpful comments.

% \section*{Acknowledgments}

% The acknowledgments should go immediately before the references.  Do
% not number the acknowledgments section. Do not include this section
% when submitting your paper for review.

\balance

\bibliography{eacl2017}
\bibliographystyle{eacl2017}

\end{document}